\documentclass[twocolumn,prb,superscriptaddress]{revtex4-1}
\usepackage{amsmath,amssymb,mathrsfs}
\usepackage{natbib}
\usepackage{subfigure}
\usepackage{tabularx}
\usepackage{epsfig}
\usepackage{longtable}
\usepackage{amsfonts}
\usepackage{rotating}
\usepackage{bbold}
\usepackage{hhline}
\usepackage{braket}
\usepackage{txfonts, comment}
\usepackage{multirow}
\usepackage{appendix}
\usepackage[unicode=true,bookmarks=true,bookmarksnumbered=false,bookmarksopen=false,breaklinks=false,pdfborder={0 0 1},backref=false,colorlinks=true]{hyperref}

\hypersetup{linkcolor=magenta,urlcolor=blue,citecolor=blue,pdfstartview={FitH},hyperfootnotes=false,unicode=true}

\def\be{\begin{equation}}
\def\ee{\end{equation}}
\def\bea{\begin{eqnarray}}
\def\eea{\end{eqnarray}}

\newcommand{\normord}[1]{:\mathrel{#1}:}

\begin{document}

\title{Random non-Hermitian action theory for stochastic quantum dynamics: 
from canonical to path integral quantization}

\author{Pei Wang}
\affiliation{Department of Physics, Zhejiang Normal University, Jinhua 321004, China}
\email{wangpei@zjnu.cn}

\begin{abstract}
We develop a theory of random non-Hermitian action that, after quantization, describes 
the stochastic nonlinear dynamics of quantum states in Hilbert space. Focusing on fermionic 
fields, we propose both canonical quantization and path integral quantization, demonstrating 
that these two approaches are equivalent. Using this formalism, we investigate the evolution 
of a single-particle Gaussian wave packet under the influence of non-Hermiticity and 
randomness. Our results show that specific types of non-Hermiticity lead to wave packet 
localization, while randomness affects the central position of the wave packet, causing the 
variance of its distribution to increase with the strength of the randomness.
\end{abstract}

\date{\today}

\maketitle

\section{Introduction}

A vector evolving in Hilbert space, governed by the deterministic linear Schr\"{o}dinger equation, 
has long been used to describe the physical states of various systems, from elementary particles 
to atomic gases and solid-state materials~\cite{Sakurai,Weinberg,Girvin}. 
In contrast, classical mechanics incorporated 
nonlinearity and randomness into the equations governing real-time dynamics---such as the 
Langevin equation or chaotic systems---much earlier~\cite{Zwanzig,Kaplan}. It was only in the 1980s that similar 
concepts gained attention in quantum mechanics~\cite{GRW,Diosi89,CSL,CSL2}. The introduction of stochastic 
nonlinear evolution in Hilbert space extended the framework of conventional quantum mechanics, with applications emerging in theories such as spontaneous wave function 
localization~\cite{GRW,Diosi89,CSL,CSL2,Penrose96,Pearle99,Bassi05,Adler07,Adler08,Bassi13} 
and the unravelling of the Lindblad equation in open quantum systems~\cite{Gisin92,Percival98,Plenio98,Daley14}.

Spontaneous wave function localization is an attempt to modify conventional quantum 
mechanics to account for the quantum measurement process. In this theory, the 
Schr\"{o}dinger equation is replaced by a stochastic nonlinear differential equation, 
which causes the wave packet of a macroscopic object to spontaneously localize in 
real space, thus explaining the emergence of pointer states in measuring 
apparatuses and random outcomes of quantum
measurement~\cite{GRW,Diosi89,CSL,CSL2,Penrose96,Pearle99,Bassi05,Adler07,Adler08,Bassi13}. 
As experimental 
efforts to test the theory have intensified in recent
years~\cite{Vinante16,Vinante17,Bahrami18,Tilloy19,Pontin19,Vinante20,Zheng20,Komori20,Donadi21,Gasbarri21,Carlesso22}, 
the theory has faced some challenges---most notably, the difficulty of incorporating symmetries such as Lorentz 
symmetry into the stochastic nonlinear equations~\cite{Myrvold17,Tumulka20,Jones20,Jones21}.

On the other hand, stochastic nonlinear dynamics of quantum states has also been 
used to describe the real-time evolution of open quantum systems through the 
so-called "unraveling" of quantum master 
equations~\cite{Gisin92,Percival98,Plenio98,Daley14,Castin95,Power96,Ates12,Hu13,Raghunandan18,Pokharel18,Weimer21,Weimer22}. 
The density matrix of open system evolves according to a quantum master equation, such as the Lindblad equation.
Since the density matrix is a vector in the extended space $\mathcal{H}^2$, solving for it is 
more challenging than solving for a pure state vector in the original Hilbert space, $\mathcal{H}$. 
A commonly used approach is to unravel the Lindblad equation into an equivalent equation 
governing the stochastic nonlinear dynamics of a state vector in $\mathcal{H}$. This method reduces 
the dimensionality from $\mathcal{H}^2$ to $\mathcal{H}$, though at the cost of turning the 
original deterministic linear equation into a stochastic nonlinear one. The well-established mathematical 
tools for linear systems do not apply in this case.

Given the challenges in solving nonlinear equations and incorporating certain symmetries, 
it remains essential to explore new formalisms for the stochastic nonlinear dynamics of 
quantum states. It has long been established that quantum models can be equivalently 
defined using the Lagrangian or its integral---the action---instead of the Hamiltonian or 
Schr\"{o}dinger equation. The path integral approach has been developed, allowing for the 
direct calculation of scattering matrices or particle propagators by integrating $e^{iS}$ 
(where $S$ is the action) over all possible paths. The action-based approach 
offers distinct advantages, particularly when dealing with symmetries~\cite{Weinberg}.
We previously developed an action-based  approach to describe stochastic dynamics 
in Hilbert space~\cite{Wang22}. This involved extending the 
concept of a definite action to a random-number-valued action, where
we constrained the action to be a real number, ensuring it remained 
Hermitian. This real-valued action forces physical states to obey a linear dynamical equation. 

In this paper, we remove this constraint and extend the 
framework to a random non-Hermitian (RNH) action, allowing the action to take arbitrary 
non-Hermitian complex values. As we will demonstrate, generalizing to non-Hermitian action introduces nonlinearity 
into the differential equation governing the physical states, transforming it into a stochastic 
nonlinear equation. On one hand, we can define a prenormalized state vector that 
still adheres to a linear dynamical equation. However, once the state vector is normalized, 
it evolves according to a nonlinear equation. The advantage of this approach lies in the fact 
that the prenormalized state vector follows a linear equation, making it much easier to solve by
using path integral techniques. Normalizing the solution at the final time 
step allows us to avoid directly solving a nonlinear equation while still capturing the 
essential nonlinear effects.

Given the advantages of easily incorporating symmetries and converting nonlinear equations 
into linear ones, we believe that the RNH action is a promising 
approach to exploring the stochastic nonlinear physics of quantum systems. In this paper, 
we demonstrate the concept of the RNH action using a model of spinless fermions in 1+1-dimensional 
spacetime. While our approach is presented in this simple context, it can be readily extended 
to higher-dimensional spacetimes, bosonic fields, or mixed systems of fermions and bosons.
We systematically show how to derive the stochastic nonlinear equation for physical states,
and prove that the path integral formulation based on the action is equivalent to the corresponding 
stochastic nonlinear equations. Furthermore, we use our formalism to study the effect of 
non-Hermiticity and randomness on particle transport.

The remainder of the paper is organized as follows: In Sec.~\ref{sec:action}, we construct 
the canonical quantization of an RNH action and derive the corresponding nonlinear 
stochastic equation for the physical state. In Sec.~\ref{sec:nonli}, we present the path 
integral approach and prove the equivalence between
canonical and path integral quantizations. Section~\ref{sec:single} focuses 
on the single-particle case, where we present the path integral technique for propagator
and provide an example illustrating how randomness and non-Hermiticity can significantly 
alter wave function behavior. Finally, in Sec.~\ref{sec:con}, we summarize our findings.

\section{Random-nonHermitian action and its Hamiltonian dynamics}
\label{sec:action}

In this paper, we consider spinless fermions in 1+1-dimensional spacetime, where 
the action is expressed as
\be
\begin{split}\label{eq:ac:ac}
S = & \int dt \int dx \
\psi^{*}(t,x) \left[ i\partial_t - \mathcal{H}_1\left(x\right) \right]\psi(t,x) \\
& - \int dW_t \int dx \
\psi^{*}(t,x) \mathcal{H}_2 \left(x\right) \psi(t,x) ,
\end{split}
\ee
where $\psi(t,x)$ is a fermionic field (valued as a Grassmann number). $\mathcal{H}_1$ and 
$\mathcal{H}_2$ are linear operators acting on the spatial coordinate $x$, such as 
$x$, $x^2$, or $\partial^2/\partial x^2$. Unlike conventional quantum field theory, 
we do not require $\mathcal{H}_1$ or $\mathcal{H}_2$ to be Hermitian operators.
$dW_t$ represents the differential of the Wiener process, a random variable with zero 
mean and variance $dt$, where $dt$ is an infinitesimal time interval. The term 
$\int dW_t$ denotes the Itô integral from stochastic calculus. Since $\mathcal{H}_2 \neq 0$, 
the action $S$ becomes a random quantity, and if either $\mathcal{H}_1$ or $\mathcal{H}_2$ 
is non-Hermitian, $S \neq S^*$, meaning the action is non-Hermitian. 

There are two main approaches to constructing a quantum theory from an action: the path 
integral approach and canonical quantization. In the following, we demonstrate 
that both approaches remain valid for our RNH action, though 
some modifications are required. Ultimately, these two methods yield equivalent quantum theories.

Let us first examine the canonical quantization method. It involves using the Legendre 
transformation to obtain the corresponding Hamiltonian, which is then used to derive 
the dynamical equation of motion for the state vector. Before applying the Legendre 
transformation, it's important to recognize that the action includes an It\^{o} integral, which cannot 
be converted into a standard time integral. One might attempt to rewrite the integral 
$\int dW_t$ as $\int dt \frac{dW_t}{dt}$; however, this is problematic because the 
Wiener process is not differentiable, meaning $\frac{dW_t}{dt}$ does not exist.
In conventional quantum theory, the action is typically expressed as the time integral of 
a Lagrangian. However, in the case of the action~\eqref{eq:ac:ac}, there is no well-defined 
Lagrangian. Instead, we define a Lagrangian integral $dS$~\cite{Wang22}, where the action 
is given by $S = \int dS$. The Lagrangian integral represents the Lagrangian's integration 
over an infinitesimal time interval. This allows us to apply the Legendre transformation to $dS$.

If we treat $\psi$ as the coordinate, the canonical conjugate momentum is $i\psi^*$. 
The presence of the $dW_t$ term does not affect this definition, as it contains no time 
derivative of $\psi$. The Legendre transformation of $dS$ gives us the Hamiltonian 
integral~\cite{Wang22}, defined as $d {H} = dt \ \int dx \ i \psi^* \partial_t \psi - dS $.
Following the canonical quantization procedure, we replace $\psi(t,x)$ with the corresponding 
operator $\hat{\psi}(t,x)$, and then transition to the Schr\"{o}dinger picture by replacing 
$\hat{\psi}(t,x)$ with $\hat{\psi}(x)$, the operator at the reference time. This transition 
has been rigorously validated in our previous work~\cite{Wang22}.
The resulting Hamiltonian integral operator is expressed as
\be\label{eq:ac:dH}
\begin{split}
d \hat{H}_t = dt \int dx \
\hat{\psi}^{\dag}(x) \mathcal{H}_1\hat{\psi}(x)  
 + dW_t \int  dx \
\hat{\psi}^{\dag}(x) \mathcal{H}_2 \hat{\psi}(x).
\end{split}
\ee
The Hamiltonian integral can be interpreted as the Hamiltonian integrated over 
an infinitesimal time interval. In the absence of randomness ($\mathcal{H}_2 = 0$), 
this reduces to the standard Hamiltonian times $dt$. Since $\mathcal{H}_1$ and $\mathcal{H}_2$ 
are non-Hermitian, $d\hat{H}_t$ becomes a RNH operator. Additionally, the Hamiltonian 
integral is time-dependent because it relies on $dW_t$, with $dW_t$ at different $t$
corresponding to independent random variables.

Before introducing the evolution of a physical state, we first need to define the prenormalized 
state, denoted as $\ket{\phi_t}$. This is a vector in Hilbert space, though its norm is not 
necessarily equal to one. The evolution of this prenormalized state vector from time 
$t$ to $t + dt$ is governed by the linear evolution operator $\hat{U}_{dt}$, such that 
$\ket{\phi_{t+dt}} = \hat{U}_{dt} \ket{\phi_t}$. The infinitesimal evolution operator $\hat{U}_{dt}$ 
is expressed as $\hat{U}_{dt} = \normord{e^{-i d\hat{H}_t}}$, where $\normord{}$ represents 
normal ordering, which places creation operators to the left of annihilation operators. 
This normal ordering is necessary to maintain consistency between the operator formalism 
and the path integral approach. In contrast to conventional quantum theory, where 
$d\hat{H}_t$ is of order $\mathcal{O}(dt)$ and terms involving the second-order 
expansion of $e^{-i d\hat{H}_t}$ can typically be neglected, here we cannot neglect 
terms of the form $\left(d\hat{H}_t\right)^2$ due to the stochastic nature of $d\hat{H}_t$ 
as given in Eq.~\eqref{eq:ac:dH}, particularly because $\left(dW_t\right)^2 = dt$. 
As a result, $\normord{e^{-i d\hat{H}_t}}$ and ${e^{-i d\hat{H}_t}}$ correspond to two 
distinct infinitesimal evolution operators in this framework.

The prenormalized state vector satisfies a linear dynamical equation. To derive this equation, 
we first decompose the Hamiltonian integral into deterministic and stochastic components, 
expressing both in terms of Hermitian operators. For this purpose, we introduce the following 
Hermitian operators:
\be
\begin{split}
\hat{H}_0 = & \ \frac{1}{2} \int dx \ \hat{\psi}^\dag(x)\left(\mathcal{H}_1+ \mathcal{H}^\dag_1\right) 
\hat{\psi}(x), \\
\hat{V}_0 =  & \ \frac{1}{2i} \int dx \ \hat{\psi}^\dag(x)\left(\mathcal{H}_1- \mathcal{H}^\dag_1\right) \hat{\psi}(x),\\
\hat{H}_R = & \ \frac{1}{2} \int dx \ \hat{\psi}^\dag(x)\left(\mathcal{H}_2+ \mathcal{H}^\dag_2\right) 
\hat{\psi}(x), \\
\hat{V}_R = & \ \frac{1}{2i} \int dx \ \hat{\psi}^\dag(x)\left(\mathcal{H}_2- \mathcal{H}^\dag_2\right) \hat{\psi}(x),
\end{split}
\ee
where $\mathcal{H}^\dag_1$ and $\mathcal{H}^\dag_2$ are the adjoint operators of $\mathcal{H}_1$ 
and $\mathcal{H}_2$, respectively. Using these Hermitian operators, we can rewrite the Hamiltonian integral as: 
$ d\hat{H}_t = \hat{H}_0 \ dt + i\hat{V}_0 \ dt + \hat{H}_R \ dW_t +  i\hat{V}_R \ dW_t$,
where $\hat{H}_0$, $\hat{V}_0$, $\hat{H}_R$, and $\hat{V}_R$ represent the standard Hermitian Hamiltonian, 
antiHermitian Hamiltonian, random Hermitian Hamiltonian, and random antiHermitian Hamiltonian, respectively.
If we consider only $\hat{H}_0$, we recover the conventional quantum theory. In this work, however,
we focus on the case where all four terms contribute. The evolution operator $\hat{U}_{dt}$ can be written
as $\hat{U}_{dt} = 1-id\hat{H}_t -\frac{1}{2}\normord{\left(d\hat{H}_t\right)^2} $. Due to the existence
of second-order terms, we introduce the quartic Hermitian operators:
\be
\begin{split}
\hat{H}_{Q} = & \int dx dy \hat{\psi}^\dag(x)\hat{\psi}^\dag(y)
\frac{\mathcal{H}_2(x)\mathcal{H}_2(y)- 
\mathcal{H}^\dag_2(x) \mathcal{H}^\dag_2(y)}{2i} \hat{\psi}(y)\hat{\psi}(x), \\
\hat{V}_Q =  & \int dx dy \hat{\psi}^\dag(x)\hat{\psi}^\dag(y)
\frac{\mathcal{H}_2(x)\mathcal{H}_2(y)+ 
\mathcal{H}^\dag_2(x) \mathcal{H}^\dag_2(y)}{-2} \hat{\psi}(y)\hat{\psi}(x).
\end{split}
\ee
With these operators, the linear differential equation for the prenormalized state vector $\ket{\phi_t}$ is given by:
\be
\begin{split}\label{eq:ac:dph}
\ket{d\phi_t} = \ & 
- i \left(\hat{H}_0+\frac{\hat{H}_Q}{2}\right) dt \ket{\phi_t}  + 
\left(\hat{V}_0+\frac{\hat{V}_Q}{2}\right) dt \ket{\phi_t} \\ &  -i \hat{H}_R \ dW_t\ket{\phi_t} +  \hat{V}_R \ dW_t \ket{\phi_t},
\end{split}
\ee
where $ \ket{d\phi_t} $ is defined as $\ket{d\phi_t} \equiv \hat{U}_{dt} \ket{\phi_{t}}-\ket{\phi_t}$.
Although the Hamiltonian integral is quadratic in nature, the second-order expansion of the evolution operator
introduces quartic terms in the dynamical equation. This leads to the emergence of the quartic Hermitian
operators $\hat{H}_Q$ and $\hat{V}_Q$, reflecting the contributions from higher-order interactions.

For an evolution from $t_0$ to $t_f$, over a finite time interval ($t=t_f-t_0$), the prenormalized state 
at the final time $t_f$ is given by
\be\label{eq:ac:phtf}
\ket{\phi_{t_f}} = \lim_{dt\to 0 } \ \normord{e^{-i d \hat{H}_{t_{N-1}}}} \ \cdots 
\ \normord{e^{-i d \hat{H}_{t_1}}} \ \normord{ e^{-i d\hat{H}_{t_0}}} \  \ket{\phi_{t_0}},
\ee
where $t= N dt$, and $t_j = j dt +t_0$ for $j=0,1,\cdots, N-1$. Note that $d\hat{H}_{t_j}$, 
evaluated at different times, represents independent random operators.
Since the evolution operator $\hat{U}_{dt}$ is nonunitary---evident from the nonHermiticity of 
$d\hat{H}_t$---it does not conserve the norm of the state $\ket{\phi_t}$. Even if we initially set 
$\braket{\phi_{t_0} | \phi_{t_0}}=1$, in general, we will find $\braket{\phi_{t_f} | \phi_{t_f}}\neq 1$ 
at the final time $t_f$. Therefore, the prenormalized state $\ket{\phi_t}$ is unsuitable for directly 
representing the physical state of the quantum system.

To represent the physical state, we must instead use the normalized state vector, defined as
\be
\ket{\bar{\phi}_t} = \frac{\ket{\phi_t}}{\sqrt{\braket{\phi_t | \phi_t}}}.
\ee
This shows that the physical state vector is simply the projection of the prenormalized vector 
onto the unit hypersphere in Hilbert space. Thus, one can calculate the prenormalized state $\ket{\phi_t}$,
which follows a linear equation and is computationally simpler, and then normalize it at the final time
to obtain the physical state. In this way, the RNH action governs a quantum dynamics that is linear for 
the prenormalized state.

On the other hand, examining the equation of motion governing the physical state reveals that it 
satisfies a truly nonlinear equation. By applying the chain rule from stochastic calculus and using
Eq.~\eqref{eq:ac:dph}, we can derive expressions for $d\braket{\phi_t | \phi_t}$ and subsequently for
$d\left(\braket{\phi_t | \phi_t}^{-1/2}\right)$. Using these, we obtain the following equation for 
the normalized physical state: 
\begin{widetext}
\be\label{eq:ac:dpt}
\begin{split}
\ket{d\bar{\phi}_t} = & \ -i dt \left(\hat{H}_0+ \frac{\hat{H}_Q}{2}\right) \ket{\bar{\phi}_t} 
+ i dt \ \langle\hat{V}_R\rangle 
\hat{H}_R \ket{\bar{\phi}_t} -i dW_t \ \hat{H}_R \ket{\bar{\phi}_t} +
 dW_t \left( \hat{V}_R - \langle \hat{V}_R\rangle\right) \ket{\bar{\phi}_t} \\ & \
+dt \left\{ \hat{V}_0- \langle
\hat{V}_0\rangle + \frac{1}{2}\left(\hat{V}_Q-\langle \hat{V}_Q\rangle\right) +
 \frac{1}{2} i \langle \left[\hat{V}_R, \hat{H}_R\right]\rangle - \frac{1}{2}\langle\hat{H}_R^2 \rangle 
 - \frac{1}{2}\langle\hat{V}_R^2 \rangle +\frac{3}{2} \langle\hat{V}_R\rangle^2 -\langle\hat{V}_R\rangle
 \hat{V}_R \right\} \ket{\bar{\phi}_t} ,
\end{split}
\ee
\end{widetext}
where $\langle\cdot\rangle$ denotes the expectation value with respect to $\ket{\bar{\phi}_t}$.
Equation~\eqref{eq:ac:dpt} is notably more complex than the standard Schr\"{o}dinger equation, 
so it warrants further explanation. The first term on the right-hand side, with the prefactor $i$, 
describes linear, unitary, and deterministic evolution, similar to the Schrödinger equation. 
Specifically, $\hat{H}_0$ represents the conventional Hamiltonian, while $\hat{H}_Q / 2$ 
introduces a quartic correction arising from the random force. The second term is also unitary 
and deterministic but introduces nonlinearity, as it depends on the expectation value 
$\langle \hat{V}_R\rangle$, which itself is a function of the physical state. The third term 
describes linear, unitary evolution influenced by randomness, as indicated by the presence 
of $dW_t$, and accounts for the effects of a stochastic force on the state vector. The fourth term 
represents non-unitary, nonlinear evolution, as it lacks the prefactor $i$ and also depends on 
$\langle\hat{V}_R\rangle$. The remaining terms in the second line of the equation account for 
nonlinear, non-unitary effects, compensating for the changes in the state vector's norm caused by 
the random terms in the first line.

Equation~\eqref{eq:ac:dpt} is too complicated to be solved directly. To determine the evolution of the
physical state, we observe that Eq.~\eqref{eq:ac:dpt} is equivalent to Eq.~\eqref{eq:ac:phtf}, which defines
a linear relationship between the initial and final states. For simplicity, let us assume both the initial and final
states are single-particle states. By inserting the single-particle completeness relation 
$\int dx \ket{x}\bra{x}\equiv 1$ into Eq.~\eqref{eq:ac:phtf}, we find $\phi(t_f,x_f) = 
\int d x_0 \braket{x_ft_f |x_0 t_0 } \phi(t_0,x_0)$, where
\be\label{eq:pa:xt}
\braket{x_ft_f |x_0 t_0 } = \bra{x_f} \normord{e^{-i d \hat{H}_{t_{N-1}}}} \cdots 
\normord{e^{-i d \hat{H}_{t_1}}} \normord{ e^{-i d\hat{H}_{t_0}}} \ket{x_0}
\ee
is the single-particle propagator, which is independent of the specific wave function $\phi$. 
Therefore, for any linear combination of initial states, the final state will be the 
corresponding linear combination of their respective final states. Notably, the normalized
physical state $\ket{\bar{\phi}_t}$ does not possess such linearity, which is evident from 
the nonlinearity of Eq.~\eqref{eq:ac:dpt}. To obtain the physical wave function at the final
time $t_f$, we can normalize $\phi(t_f,x_f)$. Hence, the computation of the final physical state
reduces to the calculation of the propagator.

\section{Path integral approach}
\label{sec:nonli}

In this section, we demonstrate that the canonical quantization, which leads to the 
propagator~\eqref{eq:pa:xt}, is equivalent to the path integral formulation by explicitly showing 
that Eq.~\eqref{eq:pa:xt} can also be derived using the path integral approach. Although we 
choose the single-particle propagator for demonstration, the generalization to the scattering 
matrix between many-body initial and final states is straightforward.
The path integral can be constructed by inserting a sequence of orthonormal basis states 
into Eq.\eqref{eq:pa:xt}. The orthonormal basis for fermionic fields has long been established 
(see, for instance, Ref.[\onlinecite{Weinberg}]), and satisfies the completeness relations 
$\int\ket{\varphi} D\varphi \bra{\varphi} \equiv 1$ and $\int \ket{\pi} D\pi \bra{\pi} \equiv 1$, 
where $\ket{\varphi}$ and $\ket{\pi}$ are the eigenstates of $\hat{\psi}(x)$ and $\hat{\psi}^\dag(x)$, 
respectively. The subsequent steps follow the same procedure as in conventional quantum field theory. 
This is because, even though $d\hat{H}_t$ is random and non-Hermitian, it remains quadratic 
in the fermionic fields.

By utilizing the completeness relation, we find
\be
\begin{split}\label{eq:p:prop}
 \braket{ x_f t_f | x_0 t_0}  \propto & \ \int  D\pi \, D\varphi \  
 \pi_{t_0}(x_0) \varphi_{t_N}(x_N)  \braket{\varphi_{t_0} | \pi_{t_0}}\\
& \times \prod^{N-1}_{j=0} \left(\braket{\varphi_{t_{j+1}}|\pi_{t_{j+1}}}
 \bra{\pi_{t_{j+1}} } \normord{e^{-id\hat{H}_{t_j}}} \ket{\varphi_{t_j}} \right),
\end{split}
\ee
where we have set $x_f \equiv x_N$ and $t_f \equiv t_N$. And $\varphi_{t_j}(x)=\varphi(t_j,x) = 
\braket{x | \varphi_{t_j}}$ and $\pi_{t_j}(x)= \pi(t_j,x) = \braket{\pi_{t_j} | x}$ are the Grassmann-valued coordinate
field and canonical conjugate momenta field, respectively~\cite{Weinberg}.
We neglect the prefactor that is independent of the initial and final states,
as it only contributes a constant to the propagator, and consequently, to the wave function.
Since the wave function must be normalized at the final time, 
considering this constant prefactor during the calculation process is unnecessary.
Using the fact that $\bra{\pi}$ and $\ket{\varphi}$ are eigenvectors
of $\hat\psi^\dag$ and $\hat\psi$, with the corresponding eigenvalues $\pi(x)$
and $-i \varphi(x)$, respectively, we obtain 
\be
\begin{split}\label{eq:p:dpp}
 \bra{\pi_{t_{j+1}} } \normord{e^{-id\hat{H}_{t_j}}} & \ket{\varphi_{t_j}}
 \\  = \braket{\pi_{t_{j+1}} | \varphi_{t_j}} & \exp\left\{ -i \left[dt \int 
 dx \left(-i \pi_{t_{j+1}}(x)\right) \mathcal{H}_1 \varphi_{t_j}(x) \right.\right.
 \\ & \left.\left. + dW_{t_j} \int dx \left(-i \pi_{t_{j+1}}(x)\right) \mathcal{H}_2 \varphi_{t_j}(x)\right]\right\},
 \end{split}
\ee
where we have used the properties of normal ordering, which are critical for
Eq.~\eqref{eq:p:dpp} to hold. Substituting Eq.~\eqref{eq:p:dpp} into Eq.~\eqref{eq:p:prop},
we find that the propagator can be expressed as
\be
\begin{split}\label{eq:p:pS}
 \braket{ x_f t_f | x_0 t_0}  \propto & \ \int  D\pi \, D\varphi \  \pi(t_0,x_0) \varphi(t_N, x_N) e^{iS},
\end{split}
\ee
where $\int D\varphi$ and $\int D\pi$ denote integration with respect to $\varphi(t_j,x_j)$ 
and $\pi(t_j,x_j)$, respectively, with $\left(t_j,x_j\right)$ running over spacetime.
The action can be written as
\be
\begin{split}\label{eq:p:act}
S = & \sum_{j=0}^{N-1} \int dx \ \pi(t_{j+1},x) 
\left[ \varphi(t_{j+1},x) - \varphi(t_j,x) \right] \\
& - \sum_{j=0}^{N-1} \int 
 dx \left(-i \pi(t_{j+1},x)\right) \left(dt \mathcal{H}_1 + dW_{t_j} \mathcal{H}_2\right) \varphi(t_j, x) \\ 
& + \int dx \ \pi(t_0,x) \varphi(t_0,x).
\end{split}
\ee
Now, let us revert to our original notation. Then, the fermionic field becomes $\varphi(t,x)
\to \psi(t,x)$. As shown in Sec.~\ref{sec:action}, the corresponding
canonical conjugate momenta is $\pi(t,x) \to i\psi^*(t,x)$. Using these notations
and taking the continuum limit $dt \to 0$, we find that the 
action~\eqref{eq:p:act} matches the original action presented in Eq.~\eqref{eq:ac:ac}, 
except for a boundary term, which can be cancelled out in the path integral approach.

Equation~\eqref{eq:p:pS} establishes the equivalence between the canonical formalism
and the path integral formalism. Due to the quadratic nature of the action, the propagator
can be explicitly determined. By discretizing the spatial coordinates, the quadratic action can
be reexpressed in matrix form as $S = \pi \ D \ \varphi = \sum_{tx, t'x'} \pi(t,x) D_{tx,t'x'} \varphi(t',x')$,
where $\pi$ and $\varphi$ are treated as row and column vectors, respectively, and
$D$ is a square matrix. Using the well-known formula for integration over Grassmann variables,
the propagator evaluates to
\be\label{eq:p:xD}
\braket{ x_f t_f | x_0 t_0} \propto D^{-1}_{t_f x_f, t_0 x_0}.
\ee
Thus, to obtain the propagator, it is sufficient to compute the inverse of $D$.

\subsection{Example}
\label{sec:mod}

We illustrate the application of Eq.~\eqref{eq:p:xD} with a specific example.
Let us choose $\mathcal{H}_2=0$, and
\be\label{eq:E:H1}
\mathcal{H}_1 = - i \left( \lambda_1 \partial_x + \lambda_2 \partial_x^2 + \cdots
+ \lambda_n \partial_x^n + \cdots \right),
\ee
where $\lambda_1, \lambda_2, \ldots$ are arbitrary complex numbers.
This choice corresponds to a deterministic non-Hermitian action.

Since this setup preserves spatial translation symmetry, the action $S$ becomes 
block-diagonal in momentum space, making the calculation of the inverse of $D$ more straightforward. 
We introduce the following Fourier transformation of the fields: $\tilde{\varphi}(t,k) = 
\int dx \, \frac{e^{-ikx}}{\sqrt{L}} \, \varphi(t,x)$ and $\tilde{\pi}(t,k) = 
\int dx \, \frac{e^{ikx}}{\sqrt{L}} \, \pi(t,x)$, where $L$ denotes the spatial length. 
Using these $k$-space fields, we can reexpress the action as
$S = \sum_k S_k$, where $S_k = \sum_{t',t} \tilde{\pi}(t',k) \tilde{D}_{t',t}(k) \tilde{\varphi}(t,k)$.
Here, $\tilde{D}(k)$ is an $(N+1)$-by-$(N+1)$ matrix. If
we arrange the row (column) indices as $(t_0, t_1, \ldots, t_N)$,
then it is written as
\be\label{eq:E:Dk}
\begin{split}
\tilde{D}(k) = \left( 
\begin{array}{ccccc}
1 & 0 & 0 & \cdots & 0 \\
-c & 1 & 0 & \cdots & 0 \\
0 & -c & 1 & \cdots & 0 \\
\vdots & \vdots & \vdots & \ddots & \vdots \\
0 & 0 & \cdots & -c & 1 
\end{array}
\right),
\end{split}
\ee
where $c = 1 - dt \left( \lambda_1(ik) + \lambda_2(ik)^2 + \cdots \right)$. Its inverse can be computed easily.
In particular, we find that $\tilde{D}^{-1}_{t_N, t_0}(k) = c^N$, which in the limit $N \to \infty$
becomes $\tilde{D}^{-1}_{t_N, t_0}(k) = e^{-t \left( \lambda_1(ik) + \lambda_2(ik)^2
+ \cdots \right)}$.

The calculation of the propagator is straightforward and is given by
\[
\braket{x_f t_f | x_0 t_0} 
\propto  \sum_k \frac{e^{ik(x_f - x_0)}}{L} \tilde{D}^{-1}_{t_N, t_0}(k).
\]
With the propagator in hand, we can now determine the final wave function at time $t_f$.
We choose the initial state to be a Gaussian wave packet with zero mean and variance $\sigma_0^2$.
The corresponding wave function is expressed as $\psi(t_0, x_0) \propto \exp
\left\{ -\frac{\left(x_0 - q_0\right)^2}{2\sigma_0^2} + ip_0 x_0 \right\}$,
where $q_0$ and $p_0$ represent the particle's initial position and momentum, respectively,
and $\sigma_0$ represents the width of the initial wave packet.
The final wave function can be obtained by integrating the product
of the initial wave function and the propagator with respect to $x_0$. For simplicity, we focus on an analytical solution,
where we retain only the first- and second-order derivatives in the Hamiltonian and set $\lambda_3 = \lambda_4 = \cdots = 0$. In this case, the final wave function becomes
\be\label{eq:E:ps}
\begin{split}
\psi(t_f, x_f) \propto \exp\left\{ -\frac{\left(
x_f - q_0 - \lambda_{1} t - i \sigma_0^2 p_0 \right)^2}{2\left(\sigma_0^2 - 2\lambda_2 t\right)} \right\}.
\end{split}
\ee

To understand the shape of the wave packet, we calculate the particle's probability density, which is the squared
absolute value of the wave function. This yields $\left| \psi(t_f, x_f) \right|^2 \propto \exp
\left\{ -\frac{(x_f - q(t))^2}{\sigma^2(t)} \right\}$, where
\be\label{eq:E:qs}
\begin{split}
q(t) = & \ q_0 + \lambda_{1R} t - \frac{2\lambda_{2I} t \left(\lambda_{1I} t + \sigma_0^2 p_0\right)}
{\sigma_0^2 - 2\lambda_{2R} t},
\\ \sigma^2(t) = & \ \sigma_0^2 - 2\lambda_{2R} t + \frac{4\lambda^2_{2I} t^2}{\sigma_0^2 - 2\lambda_{2R} t},
\end{split}
\ee
represent the central position and squared width of the wave packet, respectively. 
Here, $\lambda_{jR}$ and $\lambda_{jI}$
with $j=1,2$ represent the real and imaginary parts of the parameter $\lambda_j$.
It is evident that the wave packet retains a Gaussian shape, while its central position
and width both evolve over time, determined by the parameters $\lambda_1$ and $\lambda_2$.

For the specific case where $\lambda_2 = 0$, the packet width $\sigma$ remains constant, while
its central position $q(t)$ moves at a constant velocity $\lambda_{1R}$, as expected.
According to Eq.~\eqref{eq:E:H1}, $\mathcal{H}_1$ with $\lambda_{1R} \neq 0 $ corresponds to a
Hermitian Hamiltonian operator with a linear dispersion relation, where $\lambda_{1R}$ defines
the speed of the particle. In contrast, $\lambda_{1I} \neq 0$ corresponds to
a non-Hermitian operator, $\lambda_{1I} \partial_x$, which influences the phase of the wave function
and thus the momentum. Specifically, $\lambda_{1I} / \sigma_0^2$ determines the rate of change of momentum, 
which can be seen from exploring the phase term in Eq.~\eqref{eq:E:ps}.
However, $\lambda_{1I}$ does not directly affect the position or width of the wave packet.

The dynamics become more intriguing when $\lambda_2 \neq 0$. In this scenario, $\lambda_{2I}$
corresponds to a Hermitian Hamiltonian term, $\lambda_{2I} \partial_x^2$, which exhibits a quadratic
dispersion relation. Its impact on wave packet dynamics is well understood within the framework of conventional quantum mechanics. By setting $\lambda_1 = \lambda_{2R} = 0$
in Eq.~\eqref{eq:E:qs}, we find that a nonzero $\lambda_{2I}$ causes the wave packet center to move
with a velocity of $-2\lambda_{2I} p_0$, which is directly proportional to the initial momentum $p_0$. More important, $\lambda_{2I}$ leads to a quadratic broadening of the wave packet over time, as shown by 
$\sigma^2(t) = \sigma_0^2 + \frac{4 \lambda_{2I} t^2}{\sigma_0^2}$. 

In contrast, $\lambda_{2R}$ is associated with an anti-Hermitian Hamiltonian term, $-i\lambda_{2R} \partial_x^2$,
whose effects are less familiar in the quantum mechanics community. For $\lambda_{2R} < 0$, we can see that
it leads to a linear increase in $\sigma^2(t)$ over time, with a rate of $-2\lambda_{2R}$. 
Even when both $\lambda_{2I}$ and $\lambda_{2R}$ are nonzero, the long-time behavior of $\sigma^2(t)$
is dominated by a linear increase, which is markedly different from the quadratic broadening
caused by a Hermitian Hamiltonian term. It is well known that quadratic and linear broadening
correspond to quantum and classical transport, respectively. Thus, we can interpret the effect of the anti-Hermitian term $-i\lambda_{2R} \partial_x^2$ as inducing a transition from quantum to classical behavior.

It is also important to note that an unphysical result arises if $\lambda_{2R} > 0$. In this case,
$\sigma^2(t)$ decreases over time, eventually reaching zero at a critical time $t = t_c = 
\sigma_0^2 / (2\lambda_{2R})$, at which point the wave function becomes ill-defined. 
This outcome is due to the fact that the differential equation governing the 
wave function (Eq.~\eqref{eq:ac:dph} or~\eqref{eq:ac:dpt}), 
given a Gaussian initial condition, lacks a solution for $\lambda_{2R} > 0$ when $t \geq t_c$.

\section{Single-particle formalism}
\label{sec:single}

In this section, we focus on the evolution of a single-particle wave function. Acting with $\bra{x}$ on both sides of 
Eq.~\eqref{eq:ac:dph} and accounting for the differences between quadratic and quartic operators, we obtain
\be\label{eq:S:dp}
d_t \phi(t,x) = -i \left( \mathcal{H}_1 \, dt + \mathcal{H}_2 \, dW_t \right) \phi(t,x),
\ee
where $\phi(t,x)$ represents the prenormalized wave function, and $d_t\phi(t,x) \equiv \phi(t+dt,x) - \phi(t,x)$
represents the infinitesimal change in $\phi$ with time, while keeping $x$ fixed. Note the distinction between Eq.~\eqref{eq:S:dp} and its many-body counterpart, Eq.~\eqref{eq:ac:dph}. The quartic term in Eq.~\eqref{eq:ac:dph}, which arises from the second-order contribution of $d\hat{H}_t$, vanishes in the single-particle case. This difference exists only when the action contains the random variable $dW_t$; otherwise, the quartic term vanishes in both single-particle and many-body cases.

A standard approach for solving Eq.~\eqref{eq:S:dp} involves employing the path integral formalism, specifically a single-particle version of it. To derive this, we rewrite the differential equation~\eqref{eq:S:dp} as $\phi(t+dt,x) = e^{-id{\mathcal{H}}_t} \phi(t,x)$, where the single-particle Hamiltonian integral is expressed as
\be
d{\mathcal{H}}_t = \left( \mathcal{H}_1 + \frac{1}{2} i \mathcal{H}_2^2 \right) dt + \mathcal{H}_2 \, dW_t,
\ee
with the additional term $\frac{1}{2} i \mathcal{H}_2^2$ compensating for the second-order expansion in $e^{-id{\mathcal{H}}_t}$.

Here, $\mathcal{H}_1$ and $\mathcal{H}_2$ are composed of the position operator $\hat{x}$ and 
derivatives with respect to $x$. The derivatives can be rewritten using the single-particle momentum 
operator, defined as $\hat{p} = -i\partial_x$. In this section, we use a hat symbol to denote operators 
acting on the space of single-particle wave functions, which must be distinguished from the field operators 
acting on many-body Hilbert space in previous sections. With the momentum operator $\hat{p}$, 
we can express derivatives of any order as $\partial_x^j = \left(i\hat{p}\right)^j$. Thus, $\mathcal{H}_1$ 
and $\mathcal{H}_2$ can be written as functions of $\hat{x}$ and $\hat{p}$. If $\mathcal{H}_1$ 
and $\mathcal{H}_2$ are polynomial functions of $\hat{x}$ and $\hat{p}$, the path integral 
formulation is straightforward. In particular, if the polynomial degree is at most two, an 
analytical expression for the propagator can be derived using the path integral method.

We illustrate the method using the following example of a Hamiltonian integral:
\be\label{eq:S:mH}
d{\mathcal{H}}_t = \left( \frac{\hat{p}^2}{2m} + \lambda \hat{p} \hat{x} \right) dt + \gamma \hat{p} \, dW_t,
\ee
where $m$ is the mass of particle, while the random strength $\gamma$ and dissipation strength $\lambda$ can be any complex numbers. It is clear that $d{\mathcal{H}}_t$ is a random non-Hermitian operator. Correspondingly, we find $\mathcal{H}_1 = \left(\frac{i}{2}\gamma^2 - \frac{1}{2m}\right) \partial^2_x - i \lambda \left(\partial_x \cdot x\right)$ and $\mathcal{H}_2 = -i\gamma \partial_x$. Thus, the associated field action in Eq.~\eqref{eq:ac:ac} must be a RNH action. For a given initial wave function, the prenormalized wave function at a final time, $\phi(t_f, x_f)$, can be determined using the propagator, which is governed by $d{\mathcal{H}}_t$ and given by
\be
\braket{x_f t_f | x_0 t_0} = \bra{x_f} e^{-i d{\mathcal{H}}_{t_{N-1}}} \cdots e^{-i d{\mathcal{H}}_{t_0}} \ket{x_0}.
\ee
The path integral is performed by inserting a sequence of completeness relations. Even though the Hamiltonian integral includes a stochastic process, the path integral formalism remains essentially the same as in conventional quantum mechanics.

To demonstrate this, consider $\bra{x_{j+1}} e^{-id\mathcal{H}_{t_j}} \ket{x_j}$, where $x_j$ and $x_{j+1}$ represent the particle's position at times $t_j$ and $t_{j+1}= t_j + dt$, respectively. By inserting a momentum basis, we obtain $\int dp_j \braket{x_{j+1} | p_j} \bra{p_j} e^{-id\mathcal{H}_{t_j}} \ket{x_j}$. The key is to rewrite $\bra{p_j} e^{-id\mathcal{H}_{t_j}} \ket{x_j}$ in exponential form. Since higher-order terms like $\mathcal{O}(dt \, dW_t)$ or $\mathcal{O}(dW^3_t)$ are negligible, the exponent can be obtained by substituting $\hat{p}$ and $\hat{x}$ with $p_j$ and $x_j$, respectively, as done in conventional quantum mechanics. After integrating out $p_j$, the propagator takes the form
\be\label{eq:S:xN}
\braket{x_f t_f | x_0 t_0} \propto \int dx_1 \cdots dx_{N-1} \, e^{i\tilde{S}},
\ee
where the single-particle action is given by:
\be\label{eq:S:tS}
\tilde{S} = \sum_{j=0}^{N-1} \left\{ dt \, \frac{m}{2} \left( \frac{x_{j+1} - x_j}{dt} - \lambda x_j \right)^2 - dW_{t_j} \, \gamma m \left( \frac{x_{j+1} - x_j}{dt} - \lambda x_j \right) \right\}.
\ee
In the limit $dt \to 0$, this action becomes an integral $\tilde{S} = \int_{t_0}^{t_f} d\tilde{S}$, where 
\be
d\tilde{S} = dt \, \frac{m}{2} \left(\dot{x} - \lambda x\right)^2 - dW_t \, \gamma m \left( \dot{x} - \lambda x \right).
\ee
This expression corresponds to the Legendre transform of
$d\mathcal{H}_t$ in Eq.~\eqref{eq:S:mH}, indicating that the path integral of this model follows 
the same principles as in conventional quantum mechanics.

It is worth emphasizing that $\tilde{S}$ is a single-particle version of RNH action. Especially, it is not real when $\lambda$ takes imaginary values—in other words, it is non-Hermitian. As a result, the propagator $\braket{x_N t_N | x_0 t_0}$ loses unitarity. Moreover, $\tilde{S}$ is a random variable, whose value depends on the specific trajectory of the Wiener process. This makes the propagator itself a random variable. These characteristics are significant departures from conventional quantum mechanics, which is inherently unitary and deterministic. The lack of unitarity in the propagator does not pose any issues here, as the prenormalized wave function is not required to preserve its norm over time.

The path integral in Eq.~\eqref{eq:S:xN} is evaluated using the Gaussian integral formula. We rewrite Eq.~\eqref{eq:S:tS} in matrix form as $\tilde{S} = \frac{m}{2dt} \left( x^T A x + B^T x + C \right)$, where $x = \left(x_1, x_2, \ldots, x_{N-1}\right)^T$ represents the positions at intermediate times, $A$ is an $\left(N-1\right)$-by-$\left(N-1\right)$ matrix, $B$ is an $\left(N-1\right)$-dimensional vector, and $C$ is a constant independent of $x$. According to the Gaussian integral formula, Eq.~\eqref{eq:S:xN} yields
\be
\braket{x_N t_N | x_0 t_0} \propto \exp\left(i \frac{m}{2dt} \left(-\frac{1}{4} B^T A^{-1} B + C\right)\right).
\ee
Taking the limit $dt \to 0$, we find that the propagator becomes
\be
\braket{x_f t_f | x_0 t_0} \propto \exp\left\{ \frac{i m \lambda}{e^{2\lambda t} - 1} \left( x_f - \gamma X(t) - x_0 e^{\lambda t} \right)^2 \right\},
\ee
where $t = t_f - t_0$ represents the evolution time, and $X(t)$ is a time-dependent random variable given by $X(t) = W_t + \lambda Y_t$. Here, $W_t$ is the Wiener process, and $Y_t = \int^t_0 W_\tau e^{\lambda(t-\tau)} \, d\tau$ is an associated stochastic process. All of the randomness in the propagator is encapsulated within $X(t)$.

To understand the physics of this propagator, we consider an initial wave function that is a Gaussian wave packet centered at $q_0$, with a width $\sigma_0$ and momentum $p_0$, consistent with our assumptions in Sec.~\ref{sec:mod}. The final wave function can be again calculated using
$\psi(t_f, x_f) \propto \int \ dx_0 \ \braket{x_f t_f | x_0 t_0} \psi(t_0, x_0)$.
Through some straightforward calculations, we find that the final wave function retains its Gaussian shape, allowing its squared absolute value to be expressed as
$\left| \psi(t_f, x_f) \right|^2 \propto \exp\left\{ -{\left(x_f - q(t) \right)^2}/{\sigma^2(t)} \right\}$.
Here, the time evolution of the wave packet’s center position $q(t)$ and squared width $\sigma^2(t)$ are given by:
\be\label{eq:S:sq}
\begin{split}
\sigma^2(t) = & \ D_R + \frac{D_I^2}{D_R}, \\
q(t) = & \ N_R + \text{Re}\left[\gamma X(t)\right] + \frac{D_I}{D_R} \left( N_I + \text{Im}\left[\gamma X(t)\right] \right),
\end{split}
\ee
where the terms $D_R$, $D_I$, $N_R$, and $N_I$ are defined as follows:
\be
\begin{split}
N_R = & \ e^{\lambda_R t} \left( q_0 \cos\left(\lambda_I t\right) - p_0 \sigma_0^2 \sin\left(\lambda_I t\right) \right), \\
N_I = & \ e^{\lambda_R t} \left( q_0 \sin\left(\lambda_I t\right) + p_0 \sigma_0^2 \cos\left(\lambda_I t\right) \right), \\
D_R = & \ e^{2\lambda_R t} \left[ \cos\left(2\lambda_I t\right) \left(\sigma^2_0 + \frac{\lambda_I}{2m \left|\lambda\right|^2}\right) - \sin\left(2\lambda_I t\right) \frac{\lambda_R}{2m \left|\lambda\right|^2} \right] \\ &  - \frac{\lambda_I}{2m \left|\lambda\right|^2}, \\
D_I = & \ e^{2\lambda_R t} \left[ \cos\left(2\lambda_I t\right) \frac{\lambda_R}{2m \left|\lambda\right|^2} + \sin\left(2\lambda_I t\right) \left(\sigma^2_0 + \frac{\lambda_I}{2m \left|\lambda\right|^2}\right) \right] \\ & - \frac{\lambda_R}{2m \left|\lambda\right|^2}.
\end{split}
\ee
In these expressions, $\lambda_R$ and $\lambda_I$ denote the real and imaginary components of the dissipation strength $\lambda$, while $\gamma_R$ and $\gamma_I$ represent the real and imaginary components of the random strength $\gamma$.

The expression~\eqref{eq:S:sq} is quite complex. To better understand its implications, we analyze it step by step. First, consider the packet width, $\sigma(t)$. It is clear that $\sigma(t)$ depends solely on the dissipation parameter $\lambda$ and is independent of the random strength $\gamma$. The term $D_R$ oscillates with an amplitude proportional to $e^{2\lambda_R t}$. If $\lambda_R > 0$, there will eventually be a time when $D_R$ becomes negative. A negative $D_R$ would result in a negative $\sigma^2(t)$, which is unphysical, as discussed in Sec.~\ref{sec:mod}. In this case, the dynamical equation has no solution once $\sigma^2(t)$ reaches zero.

Thus, to ensure a physically meaningful solution without singularities over time, we require $\lambda_R < 0$. With this condition, $D_R$ converges to a finite value as $t \to \infty$, and so does $\sigma^2(t)$. The rate of convergence of the wave packet width is determined by $\left|\lambda_R\right|$. For large times, specifically $t \gg 1/\left|\lambda_R\right|$, we find that $N_R, N_I \to 0$, while $D_R$ and $D_I$ approach constant values. This leads to a stable packet width:
\be\label{eq:S:ts}
\lim_{t \to \infty} \sigma^2(t) = \frac{1}{2m(-\lambda_I)}.
\ee
Thus, only $\lambda_I < 0$ corresponds to a physically viable solution, with the long-term width of the wave packet inversely proportional to $\left|\lambda_I\right|$. This behavior differs significantly from conventional unitary quantum mechanics. In the standard Hermitian case, a Hamiltonian like $\hat{p}^2/(2m)$ results in the continuous dispersion of the wave packet, where $\sigma^2(t)$ increases as $t^2$. In contrast, the presence of a non-Hermitian Hamiltonian term, $\lambda \hat{p}\hat{x}$, along with negative $\lambda_R$ and $\lambda_I$, allows the wave packet to stabilize into a Gaussian shape with a fixed width. It’s important to note that the Hamiltonian $\lambda \hat{p} \hat{x}$ remains non-Hermitian for almost all values of $\lambda$, making this fixed-width behavior a unique non-Hermitian phenomenon. For negative values of mass $m$, a similar reasoning applies if we assume $\lambda_I$ is positive, resulting in a positive packet width.

Next, let us examine the behavior of the packet’s center position, $q(t)$. For given $\lambda_R, \lambda_I < 0$ and at sufficiently large times, i.e., $t \gg 1/\left|\lambda_R\right|$, the center position becomes:
\be
q(t) \to \text{Re}\left[\gamma X(t)\right] + \frac{\lambda_R}{\lambda_I} \text{Im}\left[\gamma X(t)\right].
\ee
Here, the center position depends on the stochastic process $X(t)$, which is itself determined by $W_t$ (the Wiener process) and $Y_t$. As a result, $q(t)$ is a random variable. To gain further insight, we can analyze the mean and variance of $q(t)$. Using properties of the Wiener process, it is straightforward to show that the mean of $q(t)$ is zero. Additionally, we find the following formulas for expectation values: $\text{E}\left[X(t)^2\right] = (e^{2\lambda t} - 1)/(2\lambda)$ and $\text{E}\left[\left|X(t)\right|^2\right] = (e^{2\lambda_R t} - 1)/(2\lambda_R)$, where $\text{E}[\cdot]$ denotes the expectation value averaged over different trajectories of the Wiener process.

Using these expressions, the variance of the center position in the limit $t\to\infty$ can be calculated as:
\be\label{eq:S:tqt}
\lim_{t \to \infty} \text{E}\left[q(t)^2\right] = -\gamma_I^2 \frac{\lambda_R}{2\lambda_I^2} - \left|\gamma\right|^2 \frac{1}{4\lambda_R} - \gamma_R \gamma_I \frac{1}{2\lambda_I}.
\ee
The steady-state variance of the center position is influenced by both the random force and the dissipation. Equation~\eqref{eq:S:tqt} shows that the variance increases with $\gamma_R$ or $\gamma_I$, reflecting the role of the random driving force. A stronger random force results in a broader distribution of the center position, pushing the particle further from the origin. When $\gamma = 0$, the random force vanishes, leading to a definite center position
(at the origin), as indicated by $\text{E}\left[q(t)^2\right] = 0$ in this case.

Furthermore, $\text{E}\left[q(t)^2\right]$ is inversely proportional to $\left|\lambda_I\right|$, similar to $\sigma^2(t)$. This implies that both the width of the wave packet and the variance of its center position increase as $\left|\lambda_I\right|$ decreases. In the limit $\left|\lambda_I\right| \to 0$, both $\sigma^2(t)$ and $\text{E}\left[q(t)^2\right]$ diverge, indicating a breakdown of the localization effect.

In the long-time limit, the wave packet tends to become localized at some position along the $x$-axis. This localization effect arises due to the negative value of $\lambda_I$, which corresponds to the non-Hermitian Hamiltonian term $i\lambda_I \hat{p}\hat{x}$. When $\gamma = 0$, meaning there is no random force, the localization position is fixed at the origin. However, for a finite random strength, the central position of the localized packet becomes a random variable with a mean of zero. The variance of this position, as given by Eq.~\eqref{eq:S:tqt}, is inversely proportional to $\left|\lambda_I\right|$. This relationship further confirms that the localization effect is driven by the negative value of $\lambda_I$.

\section{Conclusions}
\label{sec:con}

In summary, we have developed a theory of quantum dynamics governed by a random 
non-Hermitian action, extending the stochastic quantum field theory proposed in 
Ref.~[\onlinecite{Wang22}] from unitary dynamics to more general nonunitary systems. 
This work presents a canonical quantization procedure that leads to a stochastic nonlinear 
differential equation describing the evolution of the physical state in Hilbert space. 
More important, we formulated a path integral approach that is equivalent to the canonical 
quantization. Notably, the path integral method simplifies the process of obtaining the 
final-time physical state compared to solving the nonlinear differential equations in the canonical approach.

To benchmark the effects of RNH actions on particle transport, we studied the single-particle evolution, 
starting with Gaussian wave packets as initial states. For the single-particle case, the 
path integral approach was reformulated into a compact form, providing a practical tool for 
calculating the propagator. Using this propagator, we computed the prenormalized state 
based on the initial wave function, subsequently normalizing it to obtain the final-time wave 
function. We specifically focused on the effects of non-Hermiticity and randomness on the 
evolution of the wave packet.

First, we investigated the evolution under a deterministic non-Hermitian Hamiltonian defined by 
$\mathcal{H}_1 = -i\lambda_1 \partial_x - i \lambda_2 \partial_x^2$, where $\lambda_1$ 
and $\lambda_2$ are complex parameters. The absolute value of the wave function retains a 
Gaussian profile in real space, with its central position and width precisely determined. 
The linear $\partial_x$ term affects only the velocity of the central position, leaving the wave 
packet's width unchanged. In contrast, the quadratic $\partial_x^2$ term modifies the width: 
in the purely Hermitian case, the squared width $\sigma^2(t)$ grows as $t^2$, consistent with 
conventional quantum mechanics. However, when a non-Hermitian component is present, 
$\sigma^2(t)$ grows linearly with time for large times, leading to a transition from ballistic to 
diffusive behavior in the wave packet's evolution.

Next, we considered a system with a deterministic term $\mathcal{H}_1 = 
\left(\frac{i}{2}\gamma^2 - \frac{1}{2m}\right) \partial^2_x - i\lambda (\partial_x \cdot x)$, 
along with a stochastic term $\mathcal{H}_2 = -i\gamma \partial_x$, where $\lambda$ 
and $\gamma$ are complex parameters. Remarkably, when $\lambda_R, \lambda_I < 0$, 
the wave packet's width ceases to grow indefinitely and instead approaches a finite 
value in the long-time limit, inversely proportional to $|\lambda_I|$. This demonstrates 
a form of wave packet localization induced by the non-Hermitian terms, distinct from 
Anderson localization in Hermitian systems and purely a non-Hermitian effect. Additionally, 
the dissipation from $\lambda$ causes the particle to decelerate, with the central position 
of the wave packet approaching a finite value in the long run, rather than propagating 
indefinitely. The random term $\mathcal{H}_2$ introduces a stochastic force, leading to a 
steady-state position of the wave packet that becomes a random variable, whose 
distribution width increases with the strength of the randomness $\gamma$.

While this study primarily focused on single-particle dynamics, the formalism is constructed 
in a field-theoretic language, making it readily applicable to many-body systems. In future 
work, we aim to explore how the non-Hermitian and stochastic effects observed here, 
such as localization and randomness, influence the dynamics of many-body states.

\begin{acknowledgments}
The work is supported by the Junior Associates program of the Abdus Salam 
International Center for Theoretical Physics. 
\end{acknowledgments}


\end{document}